\documentstyle[11pt]{article}

\begin{document}
\begin{center}
\vspace{10mm}

{\Large Finite volume effect of nucleons and the formation of the
  quark-gluon plasma}

\vspace{12mm}
\renewcommand{\thefootnote}{\fnsymbol{footnote}}
{\normalsize
Bo-Qiang Ma$^{1}$\footnote{
Fellow of Alexander von Humboldt Foundation,
on leave from
Institute of High Energy Physics, Academia Sinica, P.O.Box 918(4),
Beijing 100039, China},
Qi-Ren Zhang$^{2}$,
D.~H.~Rischke$^{1}$, and W.~Greiner$^{1}$}

\vspace{10mm}
{\normalsize

$^{1}$  Institut f\"ur Theoretische Physik der
        Universit\"at Frankfurt am Main, Postfach 11 19 32,
        D-6000 Frankfurt, Germany}

$^{2}$  Center of Theoretical Physics, CCAST(World Laboratory),
        Beijing, China and Department
        of Technical Physics, Peking University, Beijing 100871, China

\vspace{6mm}

\end{center}

{\large \bf Abstract }
We study a thermodynamically consistent implementation of the
nucleon volume in the mean field theory, and find that this volume
has large consequences on the properties of hadronic matter
under extreme conditions such as in astrophysical objects and high
energy heavy-ion collisions.  It is shown that we can reproduce
the critical temperature $T_{c}\simeq 200$ MeV predicted by lattice
QCD calculations for the phase transition from hadronic matter
to quark-gluon plasma, by using parameters which are adjusted to fit
all empirical data for normal nuclear matter.

\vspace{10mm}
To be published in Phys.Lett.B

\newpage
One of the main goals of nuclear physics is to study the properties
of hadronic matter under extreme conditions, such as in high density
and high temperature astrophysical objects and
high energy heavy-ion
collisions \cite{GS89}, based on our knowledge of normal nuclear
matter properties.
There are some difficulties to reproduce simultaneously all
known properties for nuclear matter under normal and
extreme conditions
by using the naive conventional scenario of point-like hadrons.
It has been observed \cite{Gle86} that in the framework of
the local relativistic
mean field theory
the mass limit for neutron
stars is too low when using a nuclear equation of state
with a realistic
compression modulus $K$ around $200$ MeV.
Also, there is no reasonable phase transition to the
quark-gluon plasma for vanishing net baryon number
at large T due to the possible excitation of
a vast number of hadronic resonances
\cite{HSSG86,RGSG91}.
The purpose of this letter is to show that the
above difficulties can
be removed simultaneously by simply taking into account the effect
from the finite volumes of nucleons.
That such a consideration is reasonable
can be motivated by the success of the
Van der Waals equation of state to describe
properties of real gases.
The limits of the validity of the effective field theory of
point-like mesons and baryons are also indicated by the
off-shell behaviors of hadronic structure functions due to
the internal quark and gluon structure of hadrons \cite{Ma}.
By adopting a
thermodynamically consistent procedure in the mean field theory,
we have successfully produced \cite{ZMG92}
a mass limit for
neutron stars which is compatible with the observation, while
still kept the compression modulus for normal nuclear matter
around its empirical value $K=240$ MeV.
In this letter we will
focus our attention on the phase transition to the quark-gluon plasma
which is expected to occur in high energy heavy-ion collisions
\cite{HR80}.
We will show
that we can reproduce the critical temperature $T_{c}\simeq 200$
MeV predicted by lattice calculations
of pure SU(3) gauge theory \cite{GS89,SG86,Hwa}
using the parameters which are
adjusted to fit all empirical data for normal nuclear matter \cite{MS74},
namely, the equilibrium density
$\rho_{0}=(\frac{4\pi}{3}r^{3})^{-1}$
with $r=1.175$ fm,
the binding energy per nucleon $B/A=15.986$ MeV,
the asymmetry energy $E_{as}=36.5$ MeV, and
the compression modulus $K=240$ MeV, with the bag constant for the
quark-gluon plasma phase ranging from $B^{1/4}=125$ MeV
to about 300 MeV.

In our approach hadronic matter and the quark-gluon plasma are
described separately by different equations of state and a transition
is obtained via Gibbs' phase coexistence conditions.
The effect of the finite volume has been extensively discussed
in the ideal hadronic gas model \cite{IGM}
which cannot reproduce
the nuclear ground state properties.
We introduce the finite volume of nucleons in a
thermodynamically self-consistent
way \cite{RGSG91,ZMG92,ZL92,Zh92} in the
mean field theory framework, which has been widely  used
for the description  of nuclear matter, finite nuclei, and
nuclear dynamics \cite{SW86}.
One new aspect of our approach is that
the equation of state for the hadronic
phase is applicable to hadronic matter under extreme
conditions
as well as normal conditions, although
there are problems with
causality at very high density \cite{RGSG91,ZMG92}.
The quark-gluon plasma is treated as a gas of relativistic massless
quarks, antiquarks, and gluons, including first-order QCD corrections
and non-perturbative effects by an overall bag constant $B$.

For the sake of simplicity, we assume symmetric, homogeneous, and
isotropic
hadronic matter
with $N_{B}$ baryon species (with equal properties)
in a volume $V$,
and restrict ourselves to non-strange
degrees of freedom.
In the ideal gas model the pressures of nucleons
and antinucleons
can be written as
\begin{eqnarray}
{\cal P}^{0}_{N}(\mu,T)
=\frac{kT}{V}\zeta(\mu,T);\\
{\cal P}^{0}_{\bar{N}}(\mu,T)
=\frac{kT}{V}\zeta(-\mu,T),
\label{eq:press1}
\end{eqnarray}
with $\zeta(\mu,T)$ defined by
\begin{equation}
\frac{1}{V}\zeta(\mu,T)
=N_{B}\int\frac{2d^{3}\vec{p}}
{(2\pi)^{3}}ln[1+e^{\beta(\mu-\epsilon)}],
\label{eq:press1b}
\end{equation}
where $\beta=1/k T$ with $T$ being the temperature,
$\mu$ is the chemical potential
and $\epsilon=\sqrt{p^{2}+\chi^{2}}$ is the energy of a nucleon
of 3-momentum $\vec{p}$ and mass $\chi$.
We now consider a system of nucleons and antinucleons with
finite volume $\tau$.
Assuming that
the repulsive force due to the finite volume is operative
between nucleons or antinucleons but not between
a nucleon and an antinucleon due to the possibility of
matter-antimatter annihilation, we can simply discuss
the finite volume effect for nucleons and antinucleons
separately.
The pressures can be written as \cite{RGSG91,Zh92}
\begin{eqnarray}
{\cal P}_{N}(\mu,T)={\cal P}^{0}_{N}(\tilde{\mu},T)
=\frac{kT}{V}\zeta(\tilde{\mu},T); \\
{\cal P}_{\bar{N}}(\mu,T)={\cal P}^{0}_{\bar{N}}(\tilde{\mu}',T)
=\frac{kT}{V}\zeta(-\tilde{\mu}',T),
\label{eq:press2}
\end{eqnarray}
with the effective chemical potentials $\tilde{\mu}$ for
nucleons and
$\tilde{\mu}'$ for antinucleons given by
\begin{equation}
\tilde{\mu}=\mu-\tau{\cal P}_{N}(\mu,T);    ~~~
\tilde{\mu}'=\mu+\tau{\cal P}_{\bar{N}}(\mu,T),
\end{equation}
as required to keep thermodynamical consistency \cite{RGSG91,Zh92}.
In the mean field theory of Bodmer and Boguta \cite{BB77},
the energy density of the hadronic phase is
\begin{equation}
{\cal E}_{h}={\cal E}_{N}+{\cal E}_{\bar{N}}+
U+2\pi
\alpha_{\omega}\frac{m^{2}}{m^{2}_{\omega}}\rho_{h}^{2};
\end{equation}
with U given by
\begin{equation}
U=\frac{1}{3\pi^{2}\alpha}(1-\chi)^{2}
[1+\alpha_{1}(1-\chi)+\alpha_{2}(1-\chi)^{2}],
\end{equation}
where
$\rho_{h}$ is the net baryon
number density,
$m$ and $m_{\omega}$
are the
masses of free nucleon and $\omega$ meson,
and
$\alpha$, $\alpha_{1}$, $\alpha_{2}$, and $\alpha_{\omega}$
are the mean field theory parameters adjusted
by fitting the normal ground state properties with
the finite volume of the nucleon $\tau=\frac{4\pi}{3}a^{3}$ (with
$a=0.62$ fm) \cite{ZMG92}. The effective chemical
potentials $\nu$ for nucleons and
$\nu '$ for antinucleons are further modified to
\begin{equation}
\nu=\tilde{\mu}-
4\pi \alpha_{\omega}\frac{m^{2}}
{m^{2}_{\omega}}\rho_{h} ;
{}~~~
\nu '=\tilde{\mu}'-4\pi \alpha_{\omega}\frac{m^{2}}
{m^{2}_{\omega}}\rho_{h}
\end{equation}
due to the effects from mean fields \cite{RFSG88}.
The baryon number and kinetic energy densities for
nucleons and antinucleons are
\begin{equation}
\rho_{N}=\int\frac{2N_{B}d^{3}\vec{p}}{(2\pi)^{3}}
\frac{1-\rho_{N}\tau}
{e^{\beta(\sqrt{p^{2}+\chi^{2}}-\nu)}+1}; ~~~
\end{equation}
\begin{equation}
\rho_{\bar{N}}=\int\frac{2N_{B}d^{3}\vec{p}}{(2\pi)^{3}}
\frac{1-\rho_{\bar{N}}\tau}
{e^{\beta(\sqrt{p^{2}+\chi^{2}}+\nu ')}+1};
\end{equation}
\begin{equation}
{\cal E}_{N}=\int\frac{2N_{B}d^{3}\vec{p}}{(2\pi)^{3}}
\frac{\sqrt{p^{2}+\chi^{2}}(1-\rho_{N}\tau)}
{e^{\beta(\sqrt{p^{2}+\chi^{2}}-\nu)}+1};    ~~~
\end{equation}
\begin{equation}
{\cal E}_{\bar{N}}=\int\frac{2N_{B}d^{3}\vec{p}}{(2\pi)^{3}}
\frac{\sqrt{p^{2}+\chi^{2}}(1-\rho_{\bar{N}}\tau)}
{e^{\beta(\sqrt{p^{2}+\chi^{2}}+\nu ')}+1}.
\end{equation}
The kinetic pressures for nucleons and antinucleons can be written as
\begin{eqnarray}
{\cal P}_{N}(\mu,T)={\cal P}^{0}_{N}(\nu,T)
=\frac{kT}{V}\zeta(\nu,T); \\
{\cal P}_{\bar{N}}(\mu,T)={\cal P}^{0}_{\bar{N}}(\nu ',T)
=\frac{kT}{V}\zeta(-\nu ',T),
\label{eq:press3}
\end{eqnarray}
Thus we obtain the total pressure and the net baryon number density
for the hadronic phase:
\begin{equation}
{\cal P}_{h}(\mu,T)={\cal P}_{N}(\mu,T)+{\cal P}_{\bar{N}}(\mu,T)
-U+2\pi
\alpha_{\omega}\frac{m^{2}}{m^{2}_{\omega}}\rho_{h}^{2};
\end{equation}
\begin{equation}
\rho_{h}=\rho_{N}-\rho_{\bar{N}}.
\end{equation}
The effective nucleon mass $\chi$
is fixed by minimizing the
energy density at constant baryon number density $\rho_{h}$ and
entropy density; which is
equivalent
to maximize the pressure at fixed $T$ and $\mu$.
For the quark-gluon plasma phase,
the pressure, energy density and net baryon number density
are given
by \cite{HSSG86,S84,WRMSG89}:
\begin{eqnarray}
{\cal P}_{qgp}(\mu_{q},T_{q})
=\frac{8\pi^{2}T_{q}^{4}}{45}(1-\frac{15\alpha_{s}}{4\pi})
+N_{f}[\frac{7\pi^{2}T_{q}^{4}}{60}(1-\frac{50\alpha_{s}}{21\pi})
+(\frac{\mu_{q}^{2}T_{q}^{2}}{2}+\frac{\mu_{q}^{4}}{4\pi^{2}})
\nonumber  \\
(1-\frac{2\alpha_{s}}{\pi})]-B; ~~~~~~~~
\end{eqnarray}
\begin{equation}
{\cal E}_{qgp}
=\frac{8\pi^{2}T_{q}^{4}}{15}(1-\frac{15\alpha_{s}}{4\pi})
+N_{f}[\frac{7\pi^{2}T_{q}^{4}}{20}(1-\frac{50\alpha_{s}}{21\pi})
+3(\frac{\mu_{q}^{2}T_{q}^{2}}{2}+\frac{\mu_{q}^{4}}{4\pi^{2}})
(1-\frac{2\alpha_{s}}{\pi})]+B;
\end{equation}
\begin{equation}
\rho_{qgp}=\frac{N_{f}}{3}
(\mu_{q}T_{q}^{2}+\frac{\mu_{q}^{3}}{\pi^{2}})
(1-\frac{2\alpha_{s}}{\pi}),
\end{equation}
where $B$ is the bag constant and
$\alpha_{s}$ is the QCD running coupling constant which
depends on the quark-gluon plasma temperature $T_{q}$ and
the quark chemical potential
$\mu_{q}$ through \cite{S84}
\begin{equation}
\alpha_{s}(\mu_{q},T_{q})=\frac{4\pi}{(11-2N_{f}/3)
ln[(0.8\mu_{q}^{2}+15.622T_{q}^{2})/\Lambda^{2}]},
\end{equation}
with $\Lambda$ being the QCD scale parameter.

To examine the phase transition between hadronic matter
and the quark-gluon plasma, we apply Gibbs' conditions of phase
coexistence, i.e., ${\cal P}_{h}={\cal P}_{qgp}$;
$T=T_{q}$; and $\mu=3\mu_{q}$.
As a first approximation, we calculate the simplest
case with $N_{B}=2$ (i.e., only protons and neutrons) and
$N_{f}=2$ (i.e., only u- and d- quarks).
The range of B is chosen  from  $B^{1/4}=125$ MeV
to about 300 MeV which is consistent with the results from
a bag model analysis of hadron spectroscopy
\cite{HSSG86,IGM,WRMSG89,BM2}.
The QCD scale parameter is now quite well
determined to be $\Lambda\simeq 200$ MeV \cite{PDG88}.
Fig.~1 presents the numerical result for the $T-\mu$ phase diagram with
$B^{1/4}=200$ MeV.
In Fig.~2, the dependence of the
net baryon number densities on the temperature for the
hadronic matter and the quark-gluon plasma phase boundaries are plotted.
We see that we can reproduce a reasonable phase transition
temperature of $T_{c}=170$ MeV for vanishing net baryon number density.
$T_{c}$ changes from about 120 to 240 MeV by  changing
the parameter $B^{1/4}$ from $125$ MeV to
$300$ MeV.
Changing $a=0.62$ fm to $a=0.74$ fm (and readjusting the mean field
parameters) causes a very small change (within 1 MeV) in $T_{c}$.
By turning off the $O(\alpha_{s})$ correction
we find $T_{c}=140$ MeV, thus the first order perturbative QCD
corrections change
the ideal quark-gluon gas
result for the phase transition temperature by 20\%.
The finite volume effect is very large for
nucleons and negligible for antinucleons,
as can be seen from Fig.~1 by comparing the effective chemical
potentials $\nu$ for nucleons and $\nu'$
for antinucleons with the effective chemical potential
\begin{equation}
\tilde{\nu}=\mu-4\pi \alpha_{\omega}\frac{m^{2}}
{m^{2}_{\omega}}\rho_{h},
\end{equation}
where the contribution from the finite volume  is absent.
We see that $\nu$
is reduced by about 40\% in comparison with $\tilde{\nu}$ in the high
density region whereas $\nu'$ coincides everywhere
with $\tilde{\nu}$.

However, the finite volume effect should also play
a non-trivial role for antibaryons if there is a large number
of hadronic resonance excitations.
It has been indicated \cite{HSSG86,RGSG91}  that in a point-like hadronic
scenario
the pressure of hadronic
matter will be larger than that of the quark-gluon
plasma at large T and vanishing $\rho_{h}$ due to the
large number of hadronic resonances, thus one cannot reproduce
a phase transition to the quark-gluon plasma as predicted by
lattice QCD calculations. In our approach
this situation will not occur due to the finite volume
of baryons and antibaryons. To reflect this aspect of
the approach, in a first rough estimate
we simply  use a large $N_{B}$, which
is equivalent to assume a large number of baryons
with the same properties as nucleons. We find that
a phase
transition still occurs with $T_{c}\simeq 260$ MeV,
even if we use a very large $N_{B}$ of $2^{7}=128$.
This means that by taking into account the finite volume of baryons
we still have a
transition to the quark-gluon plasma
even if there are
a large number of hadronic resonances, thus we overcome the difficulty
of the point-like treatment of hadronic matter.
{}From the $T-\mu$ phase diagram (Fig.~3),
where the effective chemical potentials
for baryons $\nu$ and antibaryons $\nu'$ are also
plotted, we see that the finite volume effect is also large
for antibaryons at large $T$, in contrast with the simple case of only
considering
nucleons and antinucleons.  The necessity of considering other
baryonic degrees of freedom can be also seen from the large difference
between the two $T-\mu$ diagrams Figs.~1 and 3:
the effective chemical potentials
for baryons $\nu$ is much more reduced in the latter case than in the
former.

{}From Fig.~2 we see that  there are large differences between the
baryon number densities for the hadronic matter and
the quark-gluon plasma
at the phase transition at given $T$.
This aspect does not change in the case of a large baryon number
$N_{B}=128$, as can be seen from Fig.~4.
Another interesting feature
of our results is that the effective mass of the nucleon
is non-zero in the whole hadronic region:
we find $\chi \geq 0.6m$ for the case of Fig.~1 and $\chi
\geq 0.5m$ for Fig.~3.
This is different from
previous expectations of a strong reduction of
the effective baryon mass
near $T_{c}$ \cite{Theis}
but consistent with the results of ref. \cite{RGSG91}.  Of course,
the exact quantitative details depend on the specific couplings of
other baryonic resonances to the $\sigma$ field and their effective
masses. Such detailed investigations are out of the scope of the
present letter.

In summary, we have studied the finite volume effect of nucleons
by using a thermodynamically consistent procedure in the mean field
theory approach, and found that this volume effect has large
consequences in  high density and high temperature
nuclear matter concerning
the limit of neutron star
masses and the quark-gluon plasma phase transition in
high energy heavy-ion collisions.
Further studies should include a realistic mass spectrum of baryonic
resonances, instead of our simple estimate with $2^{7}$ equal
resonance species, and extend the discussion to strange degrees of
freedom.

\vspace{10mm}
We would like to acknowledge many helpful discussions with
J\"urgen Schaffner, Debades Bandyopadhyay, Raffael Mattiello, and
Horst St\"ocker.

\break
\noindent
{\large \bf Figure Captions}
\renewcommand{\theenumi}{\ Fig.~\arabic{enumi}}
\begin{enumerate}
\item  The $T-\mu$ diagram for the phase equilibrium with
$\Lambda=200$ MeV and $B^{1/4}=200$ MeV. The solid curve is
the chemical potential $\mu$. The dashed
and dot-dashed curves are the effective chemical potentials
$\nu$ for nucleons and $\tilde{\nu}$ obtained by
turning off finite volume
corrections (see eq.(24)).
The effective chemical potential $\nu'$ for antinucleons
is in coincidence with
$\tilde{\nu}$.

\item  The solid and dashed curves are the baryon number densities
$\rho_{h}$ and $\rho_{qgp}$ for hadronic matter and the
quark-gluon plasma under phase equilibrium with the same parameters
as in Fig.~1.

\item  Same as Fig.~1, but with a very large $N_{B}=128$.
The dotted
curve is the effective chemical potential $\nu'$ for
antibaryons.

\item  Same as Fig.~2, but with the same parameters as in Fig.~3.

\end{enumerate}


\newpage

\begin{thebibliography}{99}

\bibitem{GS89} See, e.g.,
               The nuclear equation of state, edited by
               W.Greiner and H.St\"ocker,
               (Plenum Press, New York,
               1989) Part A and B.

\bibitem{Gle86} N.K.Glendenning, Phys. Rev. Lett. {\bf 57} (1986) 1120;
                Nucl. Phys. {\bf A 480} (1988) 597.

\bibitem{HSSG86} U.Heinz, P.R.Subramanian, H.St\"ocker, and W.Greiner,
J. Phys. {\bf G 12} (1986) 1237.

\bibitem{RGSG91} D.H.Rischke, M.I.Gorenstein, H.St\"{o}cker,
and W.Greiner,
Z. Phys. {\bf C 51} (1991) 485.


\bibitem{Ma}  See, e.g., B.-Q.Ma, Phys. Rev. {\bf C 43} (1991) 2821;
              Int. J. Mod. Phys. {\bf E 1} (1992) 809; and
              references  therein.

\bibitem{ZMG92} Q.-R.Zhang, B.-Q.Ma, and W.Greiner,
J. Phys. {\bf G 18} (1992) 2051.

\bibitem{HR80} See, also, R.Hagedorn and J.Rafelski,
Phys. Lett. {\bf B 97} (1980) 136.

\bibitem{SG86} See, e.g., H.St\"ocker and W.Greiner,
               Phys. Rep. {\bf 137} (1986) 277.

\bibitem{Hwa} See, e.g., Quark-gluon plasma,
              edited by R.C.Hwa, (World Scientific, Singapore,
              1990).

\bibitem{MS74} W.D.Myers and W.J.Swiatecki, Ann. Phys.(NY)
{\bf 84} (1974) 186

\bibitem{IGM} See, e.g., J.Cleymans and E.Suhonen,
Z. Phys. {\bf C 37} (1987) 51;
H.Kouno and F.Takagi,
Z. Phys. {\bf C 42} (1989) 209.

\bibitem{ZL92} Q.-R.Zhang and X.-G.Li, J. Phys. {\bf G 18}(1992) L111;
               Z. Phys. {\bf A 343} (1992) 337.

\bibitem{Zh92} Q.-R.Zhang, Acta Scient. Nat. Univ. Jilin,
Suppl(Physics 1992) 1 (in Chinese).

\bibitem{BB77} J.Bodmer and A.R.Boguta, Nucl. Phys. {\bf A 292}
(1977) 413.

\bibitem{SW86} For reviews, see, e.g., B.D.Serot and J.D.Walecka, in
               Adv. Nucl. Phys., vol.16, edited by J.W.Negele and E.Vogt,
               (Plenum Press, New York, 1986);  and P.-G.Reinhard,
               Rep. Prog. Phys. {\bf 52} (1989) 439.

\bibitem{RFSG88} D.H.Rischke, B.L.Friman,
                 H.St\"ocker, and W.Greiner,
                 J. Phys. {\bf G 14}(1988) 191.

\bibitem{S84}  H.St\"ocker, Nucl. Phys. {\bf A 418}(1984)587c.

\bibitem{WRMSG89} B.M.Waldhauser, D.H.Rischke, J.A.Maruhn,
                 H.St\"ocker, and W.Greiner,
                 Z. Phys. {\bf A 43}(1989) 411.

\bibitem{BM2} Q.-R.Zhang and H.-M.Liu, Phys. Rev. {\bf C 46} (1992)
2294.

\bibitem{PDG88} See, e.g., Particle Data Group, K.Hikasa et
al., Phys. Rev. {\bf D 45} (1992) Part II, in page III. 54.

\bibitem{Theis} See, e.g., J.~Theis, G.~F.~Graebner, G.~Buchwald,
J.~A.~Maruhn, W.~Greiner, H.~St"ocker, and J.Polonyi, Phys.~Rev.~
{\bf D 28}(1983) 2286.
\end{thebibliography}
\end{document}